\documentclass[12pt,leqno]{article} \usepackage{epsfig}
\usepackage{mya4,isolatin1}

\usepackage{amsmath, amsfonts, amssymb, latexsym} 
\usepackage{myth}
\renewcommand{\vec}[1]{\underline{#1}}
\newcommand{\du}{{\protect\rule{0mm}{1mm}}\ifmmode\mathrm{d}
  \else\rmfamily{d} \fi}
\newcommand{\G}{\ifmmode {\mathcal G} \else $\mathcal G$\relax\fi}
\newcommand{\Gp}{\ifmmode {\mathcal G'} \else $\mathcal G'$\relax\fi}
\newcommand{\U}{\ensuremath{\mathcal U}}
\newcommand{\N}{\ensuremath{\mathbb N}}
\newcommand{\R}{\ensuremath{\mathbb R}}
\def\beq #1\eeq {\begin{align} #1 \end{align}}
\newcommand{\nn}{\notag}
\newcommand{\equi}{{\protect\rule{0mm}{0mm}}\ifmmode%
  \quad \Longleftrightarrow \quad \else $\Leftrightarrow$ \fi}
\newcommand{\thus}{{\protect\rule{0mm}{0mm}}\ifmmode%
  \quad \Longrightarrow \quad \else $\Rightarrow$ \fi}
\newcommand{\punkt}{~.}
\newcommand{\komma}{~,}
\newcommand{\fig}[3]{\begin{figure}[#3]
\centerline{\psfig{file=#2.eps,silent=}} 
\caption{\label{fig#2}#1}
\end{figure}}
\begin{document}
\allowdisplaybreaks
\begin{center}
\vspace*{1.0cm}

{\LARGE{\bf Dimension Theory of Graphs \hfill \\ \vspace{3mm} 
\bf and Networks}} 

\vskip 1.5cm

{\large {\bf Thomas Nowotny\quad Manfred Requardt }} 

\vskip 0.5 cm 

Institut f\"ur Theoretische Physik \\ 
Universit\"at G\"ottingen \\ 
Bunsenstrasse 9 \\ 
37073 G\"ottingen \quad Germany

\end{center}

\vspace{1 cm}

\begin{abstract}
Starting from the working hypothesis that both physics and the
corresponding mathematics have to be described by means of discrete
concepts on the Planck-scale, one of the many problems one has to face
in this enterprise is to find the discrete protoforms of the building
blocks of continuum physics and mathematics. A core
concept is the notion of {\it dimension}. In the following we develop
such a notion for irregular structures like (large) graphs and
networks and derive a number of its properties. Among other things we
show its stability under a wide class of perturbations which is
important if one has '{\it dimensional phase transitions}' in
mind. Furthermore we systematically construct graphs with almost
arbitrary '{\it fractal dimension}' which may be of some use in the
context of '{\it dimensional renormalization}' or statistical
mechanics on irregular sets.

\end{abstract} \newpage

\section{Introduction} \label{sec1}
In two recent papers (\cite{1},\cite{2}) we developed a certain
framework in form of a class of '{\it cellular network dynamics}'
which are designed to mimic the dynamics of the physical vacuum or
space-time on the Planck-scale. In doing this our working
philosophy was that both physics and the corresponding mathematics are
genuinely discrete on this primordial level. The continuum concepts of
ordinary space-time physics are then supposed to emerge from certain
discrete patterns via a kind of
'{\it renormalization group process}' on the much coarser scale of
resolution given by the  comparatively small energies of present
day high energy physics. It is one of our aims to find these discrete 
protoforms.

A crucial concepts in this context is a version of '{\it intrinsic dimension}'
of such discrete irregular networks which geometrically are
graphs. This concept should be defined in an intrinsic way, without
making 
open or implicit recourse to continuum concepts
whatsoever or kind of an embedding dimension, as we want to
understand, among other things, what properties actually are encoded in a
notion like dimension on the most fundamental physical level. On the
other 
side, we want to know how the continuum concept of dimension, which is to a
large extent of an a priori mathematical viz. geometrical origin,
comes into 
being, starting from an intrinsic property of discrete
irregular systems like e.g. general, typically very large and almost
randomly organized graphs  which are supposed to encode the `{\it
  geometrodynamics}' of space-time on Planck scale.

In section 5 of \cite{1} we introduced such a concept which seems
suitable to us and which characterizes to some extent the '{\it wiring}' of
the network. At the time of writing \cite{1} we scanned the
literature accessible to us in vain for similar ideas and got the 
impression that
such lines of thought had not been pursued in this context. Some time
later we were kindly informed by Thomas Filk that a similar
concept had been studied by himself and a couple of other physicists (see
\cite{3},\cite{4},\cite{5} and further references given there) in an 
however slightly different context. (They typically investigated the simplicial
resolution of continuous manifolds and their numerical treatment via
Monte Carlo simulations).

On the other side, at least as far as we can see, this concept had not
been systematically developed and many questions of principal interest
remained open. In the following we attempt to formulate and solve a
couple of problems which naturally emerge in this context, more
specifically we embark on developing a full fledged mathematical
machinery around this concept which then may be applied to quite
diverse fields of physics and mathematics.

Among other things we clarify the somewhat hidden relations to certain
parts of '{\it fractal geometry}' and construct graphs
with almost arbitrary '{\it fractal dimensions}' along these lines. 
Furthermore we show
that the two  at first glance almost identical definitions of
dimension we introduced in \cite{1} are actually different on certain
'{\it exceptional}' sets while, on the other side, being identical on
'{\it generic}' sets. This is a phenomenon also well known from the
various notions of dimension in fractal geometry.

While the first one, which we will call '{\it internal scaling dimension}'
in the following (it is the version which occurs under this label in
e.g. \cite{3}), appears to be more natural from a mathematical point of
view, the second one, on the other side, is in our opinion more
fundamental as far as the encoding of physical data as
e.g. the wiring of the graphs under discussion is concerned. For this reason we
call it the '{\it connectivity dimension}' as it reflects to some
extent the way the node states are interacting with each other over
larger distances via
the various bond sequences connecting them. 

Another interesting point is the structural stability of such a
concept under local and extended perturbations. We showed e.g. that if
we start from a given graph with a dimension $D$ this value remains
stable under a rather large class of bond insertions. As a consequence
one has to add bonds between increasingly distant nodes in order to
change 
the dimension of a graph. This is of some relevance if one
wants to invent dynamical mechanisms which are designed to trigger
dimensional phase transitions.

Presently we pursue several lines of research concerning
applications in quite diverse fields of physics and mathematics as
e.g.~non-commutative geometry, dimensional phase transitions (see also
\cite{2}), statistical mechanics and functional analysis. 

\section{Graph Theoretical Definitions} \label{sec2}
In this section we give the necessary definitions to define the
internal scaling dimension of graphs. Most of the notions are well 
known in graph theory but we nevertheless want to repeat them to
avoid any confusion concerning the exact definitions.

First of all we need to define an undirected simple graph. This will
be our primary object of interest.
\begin{definition}[Undirected Simple Graph] \label{def1}
  An {\em undirected simple graph} consists of two countable sets $N$
  and $B$. We denote the elements of $N$ as $n_i$ with $i \in I$, $I
  \subseteq \N$. The elements of $B$ are denoted as $b_{ik}$, $i,k \in
  I$. The set $B$ is isomorphic to a subset of $N \times N$ and the
  existence of $b_{ik}$ implies the existence of $b_{ki}$.
\end{definition}
\begin{remark}
Many mathematicians use a slightly different notation. They denote $N$ 
(nodes) as $V$ (vertices) and $B$ (bonds) as $E$ (edges).
\end{remark} 
In the following $\G=(N,B)$ will always be an undirected simple
graph. We also need the notion of the degree of a node $n_i \in N$.
\begin{definition}[Degree] \label{def2}
The {\em degree} of a node $n_i \in N$ is the number of bonds incident with
it, i.e. the number of bonds which have $n_i$ at one end. We 
count $b_{ik}$ and $b_{ki}$ only once as we interpret them as the same
bond.
\end{definition}
We assume the node degree of any node $n_i \in N$ of the graphs under
consideration to be finite.
The next step is to define a metric structure on \G. To this end we
need to define paths in \G ~and their length.
\begin{definition}[Path] \label{def3}
A {\em path} $\gamma$ of length $l$ in \G ~is an ordered $(l+1)$ tuple
of nodes $n_i \in N$, $i \in I$, $I=\{0,\dots,l\}$ with the properties
$n_{i+1} \neq n_{i}$ and $b_{i \, i+1} \in B$.
\end{definition}
\begin{remark}
A single node $n_i \in N$ is a path of length $0$.
\end{remark}
This definition encodes the obvious idea of a path in \G ~allowing
multiple transversals of nodes or bonds.  Jumps across non-existent
bonds and stays at a single node are not allowed. Sometimes this
notion of a path is also called a {\em bond sequence}.

Slightly different definitions are also quite common. The path
often is restricted to contain any bond in $B$ at most once. Sometimes
even the repetition of nodes in a path is excluded. We will call a path
with this property -- that all $n_i \in \gamma$ are pairwise different -- a
simple path.

The concept of paths on \G ~now leads to a natural definition for the
distance of two nodes $n_i$ and $n_j \in N$, namely the length of the
shortest path connecting $n_i$ and $n_j$. 
\begin{definition}[Metric] \label{def4}
A {\em metric} $d$ on \G ~is defined by
\begin{align}
d(n_i,n_j) := \left\{ \begin{array}{lc} {\min}\{l(\gamma): n_i, n_j \in
    \gamma \} & \mbox{if such $\gamma$ exist} \\ \infty &
    \mbox{otherwise,}
\end{array} \right.
\end{align}
in which $l(\gamma)$ denotes the length of $\gamma$.
\end{definition}
That this actually defines a metric is easily established. Finally we
need the notion of neighborhoods which follows canonically from the
metric.
\begin{definition}[Neighborhood] \label{def5}
Let $n_i \in N$ be an arbitrary node in \G. An $n$- neighborhood of
$n_i$ is the set $\U_n(n_i):=\{n_j \in N: d(n_i, n_j) \leq n\}$.
\end{definition}
\begin{remark}
The topology generated by the $n$-neighborhoods is the discrete
topology as should be expected from the construction and the
discreteness of graphs.
\end{remark}
We will denote the {\em surface} or {\em boundary} of the neighborhood
$\U_n(n_i)$ as $\partial\U_n(n_i)
:= \U_n(n_i) ~\backslash ~\U_{n-1}(n_i)$, $\partial \U_0(n_i)= \{n_i\}$
and the cardinality of $\U_n(n_i)$ and $\partial \U_n(n_i)$ as
$|\U_n(n_i)|$ and $|\partial \U_n(n_i)|$ respectively.

\section{Dimensions of Graphs and Networks} \label{sec3}
Now we have all the tools to define the central notion of this paper,
the notion of the {\em internal scaling dimension} of \G.
\begin{definition}[Internal Scaling Dimension] \label{def6}
  Let $x \in N$ be an arbitrary node of ~\G. Consider the sequence of
  real numbers $D_n(x):= \frac{\ln|\U_n(x)|}{\ln(n)}$. We say
  $\underline{D}_S(x):= \liminf_{n \rightarrow \infty} D_n(x)$ is the
  {\em lower} and $\overline{D}_S(x):= \limsup_{n \rightarrow \infty}
  D_n(x)$ the {\em upper internal scaling dimension} of \G ~{\em
  starting from $x$}. If $\underline{D}_S(x)= \overline{D}_S(x)=: D_S(x)$ we
  say {\G ~has internal scaling dimension $D_S(x)$ starting from
  $x$}. Finally, if $D_S(x)= D_S$ $\forall x$, we simply say \G ~has
  {\em internal scaling dimension $D_S$}.
\end{definition}
A second notion of dimension we want to introduce is the {\em connectivity
dimension} which is based on the surfaces of neighborhoods 
$\partial\U_n(n_i)$ rather than on the whole neighborhoods $\U_n(n_i)$.
\begin{definition}[Connectivity Dimension] \label{def7}
Let $x \in N$ again be an arbitrary node of \G. We set $\tilde{D}_n(x)
:= \frac{\ln|\partial \U_n(x)|}{\ln(n)} +1$ and define
$\underline{D}_C(x) := \liminf_{n \rightarrow \infty} \tilde{D}_n(x)$ 
as the {\em lower} and $\overline{D}_C(x) := \limsup_{n \rightarrow 
\infty} \tilde{D}_n(x)$ as the {\em upper connectivity dimension}. 
If lower and upper dimension
coincide, we say \G ~has {\em connectivity dimension} $D_C(x) := 
\overline{D}_C(x) = \underline{D}_C(x)$ {\em starting from}
$x$. If $D_C(x) = D_C$ for all $x \in N$ we call $D_C$
simply the {\em connectivity dimension} of \G.
\end{definition}
One could easily think that both notions of dimension are equivalent.
This is however not the case as one definition is stronger
than the other which will be shown in detail in \ref{sec3.2}.

The internal scaling dimension is rather a mathematical concept
and is related to well known dimensional concepts in fractal 
geometry as we will see in \ref{sec4.2}. The connectivity dimension
on the other hand seems to be a more physical concept as it measures
more precisely how the graph is connected and thus how nodes can
influence each other.

In the following section we want to establish the basic properties
of the internal scaling dimension of graphs.

\subsection{Basic Properties of the Internal Scaling Dimension}
\label{sec3.1} 
The first lemma gives us a criterion for the uniform
convergence of $\underline{D}_S(x)$ or $\overline{D}_S(x)$ to some common 
$\underline{D}_S$ or $\overline{D}_S$ for all nodes $x$ in \G.
\begin{lemma} \label{lem2}
Let $x$,$y \in N$ be two arbitrary nodes in \G ~with $d(x,y) < \infty$. 
Then $\underline{D}_S(y)=\underline{D}_S(x)$ and $\overline{D}_S(y)=
\overline{D}_S(x)$. 
\end{lemma}
\begin{proof}
Let $a:=d(x,y)$ be the distance of the nodes $x$ and $y$. We have
\begin{align}
& \U_{n-a}(y) \subseteq \U_n(x) \subseteq \U_{n+a}(y) \\ 
\thus & \frac{\ln|\U_{n-a}(y)|}{\ln(n)} \leq \frac{\ln|\U_n(x)|}{\ln(n)} 
\leq \frac{\ln|\U_{n+a}(y)|}{\ln(n)} \label{eqn5} \\
\thus & \frac{\ln|\U_{n-a}(y)|}{\ln(n-a) + \ln\big(\frac{n}{n-a}\big)} \leq 
\frac{\ln|\U_n(x)|}{\ln(n)}
\leq \frac{\ln|\U_{n+a}(y)|}{\ln(n+a)-\ln\big(\frac{n+a}{n}\big)} \\
\thus & \underline{D}_S(x) = \liminf_{n \rightarrow \infty} \frac{\ln|\U_n(x)|}
{\ln(n)} = \liminf_{n \rightarrow \infty} \frac{\ln|\U_n(y)|}{\ln(n)} 
= \underline{D}_S(y) \punkt
\end{align}
Similarly we get $\overline{D}_S(x)= \overline{D}_S(y)$.
\end{proof}

Another rather technical lemma provides us with a convenient method to
calculate the dimension of certain graphs, e.g. the self-similar or 
hierarchical graphs we construct in \ref{sec4.2}. It shows that under one
technical assumption the convergence of a subsequence of $D_n(x)$ is
sufficient for the convergence of $D_n(x)$ itself. 
\begin{lemma} \label{lem3}
Let $x \in N$ be an arbitrary node of \G ~and let $(|\U_{n_k}(x)|)_{k
  \in \N}$ be a subsequence of $(|\U_n(x)|)_{n \in \N}$. 
  There may exist a number $1 >
  c > 0$ such that $\frac{n_k}{n_{k+1}} \geq c$ holds for 
all $k \geq K \in \N$. 
Then $\liminf_{k \rightarrow \infty}
\frac{\ln|\U_{n_k}(x)|}{ln(n_k)} = \liminf_{n \rightarrow \infty} D_{n}(x)=
\underline{D}_S(x)$ 
and similar for $\overline{D}_S(x)$.
\end{lemma}
\begin{proof}
Let $n \in \N$ be an arbitrary natural number. We find a $k \in \N$
such that $n_k \leq n \leq n_{k+1}$. As the sequence $(|\U_n(x)|)$
is monotone this implies $|\U_{n_k}(x)| \leq |\U_n(x)| \leq
|\U_{n_{k+1}}(x)|$. Therefore we get
\begin{align}
& \frac{\ln|\U_{n_k}(x)|}{\ln(n)} \leq \frac{\ln|\U_n(x)|}{\ln(n)}
\leq \frac{\ln|\U_{n_{k+1}}(x)|}{\ln(n)} \\
\thus & \frac{\ln|\U_{n_k}(x)|}{\ln(n_k) + \ln\big(\frac{n}{n_k}\big)} 
\leq \frac{\ln|\U_n(x)|}{\ln(n)}
\leq \frac{\ln|\U_{n_{k+1}}(x)|}{\ln(n_{k+1}) + \ln\big(\frac{n}{n_{k+1}}
\big)} \\
\thus & \frac{\ln|\U_{n_k}(x)|}{\ln(n_k) + \ln(\frac{1}{c})}
\leq \frac{\ln|\U_n(x)|}{\ln(n)}
\leq \frac{\ln|\U_{n_{k+1}}(x)|}{\ln(n_{k+1}) + \ln(c)} \\
\thus & \liminf_{n \rightarrow \infty} D_{n}(x) =
\liminf_{k \rightarrow \infty} \frac{\ln|\U_{n_k}(x)|}{\ln(n_k)} \punkt
\end{align}
The same proof holds for $\limsup$.
\end{proof}
This result is well known in the context of calculation schemes for 
dimensions in fractal geometry, see e.g. \cite{6}.

Naturally one also may ask how the internal scaling dimension
behaves under insertion of bonds into \G. We were able to show that it
is pretty much stable under any local changes. We state this in the
following lemma.
\begin{lemma} \label{lem4}
Let $k \in \N$ be
a positive natural number and $x \in N$ a node in \G. 
Insertion of bonds between arbitrary many pairs of nodes ($y$, $z$)
obeying the relation $d(y,z) \leq k$ does not change  
$\underline{D}_S(x)$ or $\overline{D}_S(x)$.
\end{lemma}
\begin{proof}
We denote the new graph built by insertion of new bonds into \G ~as \Gp
~and accordingly the neighborhoods in \Gp ~as $\U'_n(\cdot)$. Being a node
in \G, $x$ is also a node in \Gp. The restriction on the choice of
additional bonds in \Gp ~implies that even if we connect every node $y \in N$
with every node in $\U_k(y)$, which is the maximum we are allowed to do,
we still can't get beyond $\U_n(x)$ with less or equal 
$\lfloor \frac{n}{k} \rfloor$ steps,
\begin{align}
  & \U_{\lfloor\frac{n}{k}\rfloor}(x) \subseteq 
  \U'_{\lfloor\frac{n}{k}\rfloor}(x) \subseteq \U_n(x) \\
  \thus &
  \frac{\ln|\U_{\lfloor\frac{n}{k}\rfloor}(x)|}{\ln(\lfloor\frac{n}{k}\rfloor)}
  \leq
  \frac{\ln|\U'_{\lfloor\frac{n}{k}\rfloor}(x)|}
  {\ln(\lfloor\frac{n}{k}\rfloor)}
  \leq \frac{\ln|\U_n(x)|}{\ln(\lfloor\frac{n}{k}\rfloor)} \punkt
\end{align}
Because $\lfloor \frac{n}{k} \rfloor \geq \frac{n}{2k}$ for sufficiently large
$n$, we immediately get
\begin{align}
  \frac{\ln|\U_{\lfloor\frac{n}{k}\rfloor}(x)|}{\ln(\lfloor\frac{n}{k}\rfloor)}
  \leq
  \frac{\ln|\U'_{\lfloor\frac{n}{k}\rfloor}(x)|}
  {\ln(\lfloor\frac{n}{k}\rfloor)}
  \leq \frac{\ln|\U_n(x)|}{\ln(n) - \ln(2k)} \\
\thus \liminf_{n \rightarrow \infty} \frac{\ln|\U'_n(x)|}{\ln(n)}
= \liminf_{n \rightarrow \infty} \frac{\ln|U_n(x)|}{\ln(n)}
\end{align}
where in the last step lemma \ref{lem3} has been used.
The identical result holds for $\limsup$.
\end{proof}
\begin{remark}
  Obviously the insertion of a {\em finite} number of additional bonds
  between nodes $y$ and $z$ with $d(y,z) < \infty$ doesn't change the
  internal scaling dimension either. Therefore we can slightly
  generalize lemma \ref{lem4} by changing our requirements to the
  following. Only bonds between nodes of finite distance and
  only finitely many bonds between nodes of distance $d(y,z) > k$ are
  inserted into \G ~to form \Gp. Then \Gp ~still has the same
  internal scaling dimensions $\underline{D}_S$ and $\overline{D}_S$
  as \G.
\end{remark}
\paragraph{Conclusions.}
We have seen that the internal scaling dimension does not depend on
the node from which we start our calculation and that under not too
strong conditions even the convergence of a subsequence of the relevant
sequence $D_n(x)$ is sufficient to calculate $\underline{D}_S$ and 
$\overline{D}_S$.
Furthermore the dimension is stable under local changes in the wiring
of the graph. This is a very desirable feature for physical reasons.
Furthermore it shows that a mechanism inducing dimensional phase 
transitions has to relate nodes of increasing distance, i.e.
has to change the graph non-locally. We will illustrate this fact with an
example in \ref{sec4.2.5}.

\subsection{Relations Between Internal Scaling Dimension and Connectivity
Dimension} \label{sec3.2}
As already stated above the two concepts of dimension we introduced are
not equivalent. In the following lemma we show that the existence of the
connectivity dimension implies the existence of the internal scaling
dimension and that they then have the same value.
\begin{lemma} \label{lem1}
  Let $x \in N$ again be an arbitrary node in \G. In the case that the
  limit ~$\lim_{n \rightarrow \infty} \frac{\ln|\partial
  \U_n(x)|}{\ln(n)} =: D_C(x)-1$ exists with $D_C(x) > 1$, 
  \G ~has internal scaling
  dimension $D_S(x)=D_C(x)$ starting from $x$.
\end{lemma}
\begin{proof}
We know that $D_C(x) > 1$ exists and have to show that
  this implies the existence of $\lim_{n \rightarrow \infty}
  \frac{\ln|\U_n(x)|}{\ln(n)}$ and that the limit is $D_C(x)$. Let
  $D:= D_C(x)$ and
  $\epsilon > 0$ be an arbitrary positive number small enough such
  that $D-1-\epsilon >0$. From the convergence of
  $\frac{\ln|\partial \U_n(x)|}{\ln(n)}$ we know that we can find 
  $N \in \N$ such that
\begin{align}
& \left| \frac{\ln|\partial \U_n(x)|}{\ln(n)} - D +1 \right| < \epsilon
\qquad \forall n \geq N \\ \thus & - \epsilon < \frac{\ln|\partial
\U_n(x)|}{\ln(n)} - D + 1 < \epsilon \\ \thus & (D-1- \epsilon ) \ln(n)
< \ln|\partial \U_n(x)| < (D-1 + \epsilon ) \ln(n) \\ \thus & n^{D-1 -
\epsilon} < |\partial \U_n(x)| < n^{D-1 + \epsilon} \punkt
\end{align}
On the other hand we naturally have
\begin{align}
& |\U_n(x)| = \sum_{j= 0}^n |\partial \U_j(x)| \\ 
 \thus
& K(N) + \sum_{j=N+1}^n j^{D-1-\epsilon} \leq |\U_n(x)| \leq K(N)
+ \sum_{j=N+1}^n j^{D-1+\epsilon}
\end{align}
in which $K(N)=\sum_{j=0}^N |\partial\U_j(x)|$.
Now we can give a lower bound for the sum on the left hand side and an
upper bound for the one on the right hand side by replacing them with
integrals.
\begin{align}
\sum_{j=N+1}^n j^{D-1-\epsilon } & \geq \int_N^n j^{D-1-\epsilon } \du
j = \frac{j^{D-\epsilon}}{D-\epsilon } \bigg|_N^n \\
\sum_{j=N+1}^n
j^{D-1+\epsilon } & \leq \int_{N+1}^{n+1} j^{D-1+\epsilon} \du j =
\frac{j^{D+\epsilon}}{D+\epsilon } \bigg|_{N+1}^{n+1}
\end{align}
With these bounds we get
\begin{align}
  & \ln \left(K(N) +  \textstyle\frac{n^{D-\epsilon } - N^{D-\epsilon}}
    {D-\epsilon } \right) \leq \ln|\U_n| \leq \ln
    \left(K(N)+ \textstyle\frac{(n+1)^{D+\epsilon } - (N+1)^{D+\epsilon}}
    {D+\epsilon } \right)  \\
&  \thus  \ln(n^{D-\epsilon }) +
  \ln\left( \textstyle\frac{K(N)}{n^{D-\epsilon}} +  \textstyle\frac{1}
    {D-\epsilon} \left(1- \textstyle\frac{N^{D-\epsilon}}
      {n^{D-\epsilon}}\right)\right)
  \leq \ln|\U_n| \\
    & \hphantom{\thus  \ln(n^{D-\epsilon })} 
    \leq \ln\left((n+1)^{D+\epsilon }\right)+
    \ln\left( \textstyle\frac{K(N)}{(n+1)^{D+\epsilon}} + 
      \textstyle\frac{1}{D+\epsilon}
      \left(1- \textstyle\frac{(N+1)^{D+\epsilon}}
        {(n+1)^{D+\epsilon}}\right) \right) \nn
\end{align}
Because the arguments of the second logarithm on each side are uniformly 
boun\-ded for any $n \in \N$ and $\lim_{n \rightarrow \infty} 
\frac{\ln(n+1)}{\ln(n)} = 1$,
we can find an $N' \in \N$, $N' \geq N$ such that $\forall n \geq N'$
\begin{align}
  & D-\epsilon + \frac{\ln\left(\frac{K(N)}{n^{D-\epsilon}} - \frac{1}
    {D-\epsilon}
    \left(1-\frac{N^{D-\epsilon}}{n^{D-\epsilon}}\right)\right)}{\ln(n)}
 \geq  ~D - 2 \epsilon \quad \text{and} \\
 & (D+\epsilon) \frac{\ln(n+1)}{\ln(n)} +
        \frac{\ln\left(\frac{K(N)}{(n+1)^{D+\epsilon}} +
            \frac{1}{D+\epsilon} \left(1-\frac{(N+1)^{D+\epsilon}}
              {(n+1)^{D+\epsilon}}\right)\right)}{\ln(n)} 
        \leq  ~D + 2 \epsilon \punkt 
\end{align}
From this we immediately find
\begin{align}
\left| \frac{\ln|\U_n|}{\ln(n)} -D \right| \leq 2 \epsilon \quad
\forall n \geq N' \punkt
\end{align}
\end{proof}
Inversely, the existence of the internal scaling dimension does not
imply the existence of the connectivity dimension. We illustrate
this fact with the following example.
\fig{Example of a graph with strange behavior of $\tilde{D}_n(x_0)
= \frac{\ln|\partial\U_n(x_0)|}{\ln(n)}$}{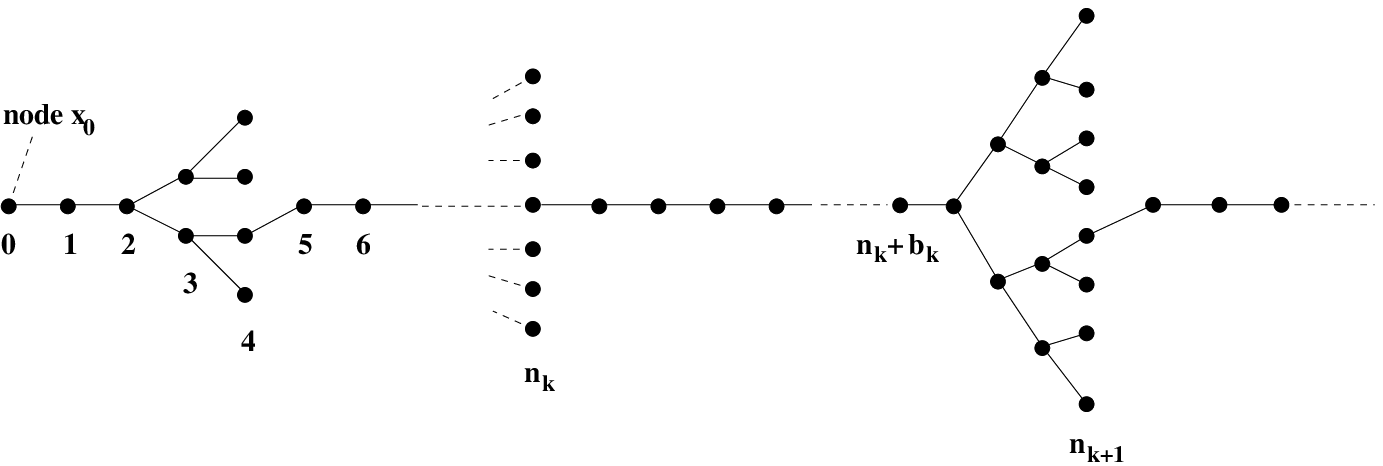}{tbp}
\begin{example}
  We will construct a graph \G ~with uniformly bounded node degree,
  degree of $x \in N$ less or equal $d \geq 3$, which has internal
  scaling dimension $D_S= D>1$ but the connectivity dimension 
  $\lim_{n \rightarrow
  \infty}\frac{\ln|\partial \U_n(x_0)|}{\ln(n)}$ does not exist and even
  $\limsup_{n \rightarrow \infty}\frac{\ln|\partial \U_n(x_0)|}{\ln(n)} = D
  \not= D-1$, i.e.~$\overline{D}_C(x_0) = \overline{D}_S(x_0)+1$. 
  To this end we construct a ``linear graph'' in the
  fashion depicted in figure \ref{figtree}. In the figure $d$ is equal
  to $3$. The main idea of the construction is to let $|\partial \U_n(x_0)|$
  oscillate so much that $\lim_{n \rightarrow \infty} \tilde{D}_n(x_0)$
  does not exist any more but we still can have convergence of $D_n(x_0)$
  and thus the internal scaling dimension exists.

  We choose the numbers $n_k$ such that $n_{k+1} \!=\! c \,n_k$ with some
  $c>0$. For technical reasons we choose $c > d^{1/D}$.
  With this
  choice we already fulfill the prerequisite to use lemma
  \ref{lem3}.
  
  Let us denote the ``leftmost'' node as $x_0$. All distances will
  refer to $x_0$ as the origin. The construction is determined by the
  following requirements. From distance $n_k$ to $n_k + b_k$ the graph
  is a simple string of nodes and from distance $n_k + b_k + 1$ to
  $n_{k+1}$ a complete\footnote{In a complete tree graph every node has 
maximal degree.}
  $(d-1)$-nary\footnote{In a $(d-1)$-nary tree graph every node has
$(d-1)$ or less children such that the degree of each node is bounded by $d$.}
  tree graph. $b_k$ is chosen to
  be $b_k = \max\{b \in \{0, \dots, n_{k+1}-n_k\} : |\U_{n_{k+1}}| \geq
  (n_{k+1})^D\}$. This means that we start the $(d-1)$-nary tree as late as
  possible to still be sure to surpass our aim of $|\U_{n_{k+1}}| =
  (n_{k+1})^D$. It is easily established that $n_{k+1} - n_k$ gets
  large enough for $n_k \geq N$ with some $N \in \N$ to contain the
  necessary $(d-1)$-nary tree. A necessary and sufficient condition for
  this is
\begin{align}
  (d-1)^{n_{k+1}-n_k} \geq n_{k+1}^D - n_k^D \\ \iff (d-1)^{c n_k
  -n_k} \geq c^D n_k^D - n_k^D \\ \iff (d-1)^{n_k(c-1)} \geq (c^D -1)
  n_k^D
\end{align}
which certainly holds for any $n_k \geq N$ with sufficiently large $N
\in \N$ because the exponential function grows faster than any polynomial.
The part of the graph where $n_{k+1}- n_k$ might be to small for the
above construction, we choose to be of arbitrary form with $|\U_{n_k}|
= \lfloor n_k^D \rfloor$.

Now we calculate the internal scaling dimension of the constructed
graph.  We know $\forall n_k \geq N$ 
\begin{align}
  \frac{\ln|\U_{n_k}(x_0)|}{\ln(n_k)} = \frac{\ln(n_k^D +
    \Delta_k)}{\ln(n_k)} \komma
\end{align}
where $\Delta_k$ is the additional number of nodes we get because of
the usage of {\em complete} tree graphs. From the construction
principle we know 
\begin{align}
  \Delta_k \leq |\partial \U_{n_k}(x_0)| \leq (d-1) |\partial \U_{{n_k}-1}(x_0)|
  \leq (d-1) |\U_{n_k-1}(x_0)| \leq (d-1) n_k^D \komma
\end{align}
which is a rather crude estimate. Nonetheless we get
\begin{align}
  & \frac{\ln(n_k^D)}{\ln(n_k)} \leq \frac{\ln|\U_{n_k}(x_0)|}{\ln(n_k)}
  \leq \frac{\ln(d n_k^D)}{\ln(n_k)} \\ 
  & \thus \lim_{k \rightarrow \infty} \frac{\ln|\U_{n_k}(x_0)|}{\ln(n_k)} 
  = D \punkt
\end{align}
Using lemma \ref{lem3} we get
\begin{align}
  D_S(x_0) = \lim_{n \rightarrow \infty} \frac{\ln|\U_n(x_0)|}{\ln(n)} = D \punkt
\end{align}
Finally we apply lemma \ref{lem2} and get the dimension $D$ starting from any
node.

On the other hand we have to consider $\liminf$ and $\limsup$ of the
sequence $\frac{\ln|\partial \U_n(x_0)|}{\ln(n)}$. The $\liminf$ is
trivial because $|\partial \U_{n_k+1}(x_0)| = 1$ which implies that $\liminf_{n
\rightarrow \infty} \frac{\ln|\partial \U_n(x_0)|}{\ln(n)} = 0$. As far
as the $\limsup$ is concerned we know
\begin{align}
  |\U_{n_{k+1}}(x_0)| - |\U_{n_k}(x_0)| =  b_k +
  \sum_{j=0}^{a_k} (d-1)^j =  b_k + \frac{(d-1)^{a_k+1}-1}{d-2} \label{eqn10}
\end{align}
with  $a_k= n_{k+1} -(n_k +b_k)$.
On the other hand
\begin{align}
  & |\U_{n_{k+1}}(x_0)| - |\U_{n_k}(x_0)| = n_{k+1}^D + \Delta_{k+1} -
  (n_k^D + \Delta_k) \label{eqn11} \punkt
\end{align}
Using (\ref{eqn10}), (\ref{eqn11}), ~$\Delta_k \leq (d-1) n_k^D$, ~ 
$b_k \leq n_{k+1} - n_k$, ~$c > d^{1/D}$~ and ~$|\partial \U_{n_{k+1}}|(x_0)=
 (d-1)^{a_k}$, we get after a short calculation that
\begin{align} 
 & D + \frac{\ln
  \left(\frac{1}{d-1}-\frac{d-2}{d-1}(1-\frac{1}{c})
  n_{k+1}^{1-D}\right)}{\ln(n_{k+1})} \leq
  \frac{\ln|\partial \U_{n_{k+1}}(x_0)|}{\ln(n_{k+1})} \\ 
  \thus & \limsup_{k
  \rightarrow \infty} \frac{\ln|\partial \U_n(x_0)|}{\ln(n)}
  \geq D \punkt \label{eqn1}
\end{align}
But we always have
\begin{align}
&  \frac{\ln|\partial \U_n(x_0)|}{\ln(n)} \leq
  \frac{\ln|\U_n(x_0)|}{\ln(n)} \\ 
\thus  & \limsup_{n \rightarrow
  \infty}\frac{\ln|\partial \U_n(x_0)|}{\ln(n)} \leq D \punkt
\end{align}
Taking this together with (\ref{eqn1}) we finally get
\begin{align}
  \limsup_{n \rightarrow \infty}\frac{\ln|\partial
    \U_n(x_0)|}{\ln(n)} = D \punkt
\end{align}
\end{example}
This example shows that we can't get much information about the
behavior of $|\partial \U_n(x_0)|$ from the existence and value of the
internal scaling dimension $D_S$ of \G. The only always valid
assertion is $\limsup_{n \rightarrow \infty} \frac{\ln|\partial
\U_n(x)|}{\ln(n)} \leq D_S(x)$ $\forall x \in N$.

\section{Construction of Graphs} \label{sec4}
In the following we want to show how to construct graphs of arbitrary
real internal scaling dimension. We also want to investigate the
connections between the internal scaling dimension of graphs and the
box counting dimension of fractal sets. As will been seen below there
is a strong relationship between self similar sets and what we also
want to call self similar graphs with non-integer internal scaling
dimension.

\subsection{Conical Graphs with Arbitrary Dimension} \label{sec4.1}
For the sake of simplicity we concentrate our discussion on graphs
with dimension $1 \leq D \leq 2$. Graphs with higher dimension are
easily constructed using a nearly identical scheme.

\fig{Example of a $\frac{5}{3}$ dimensional conical graph}{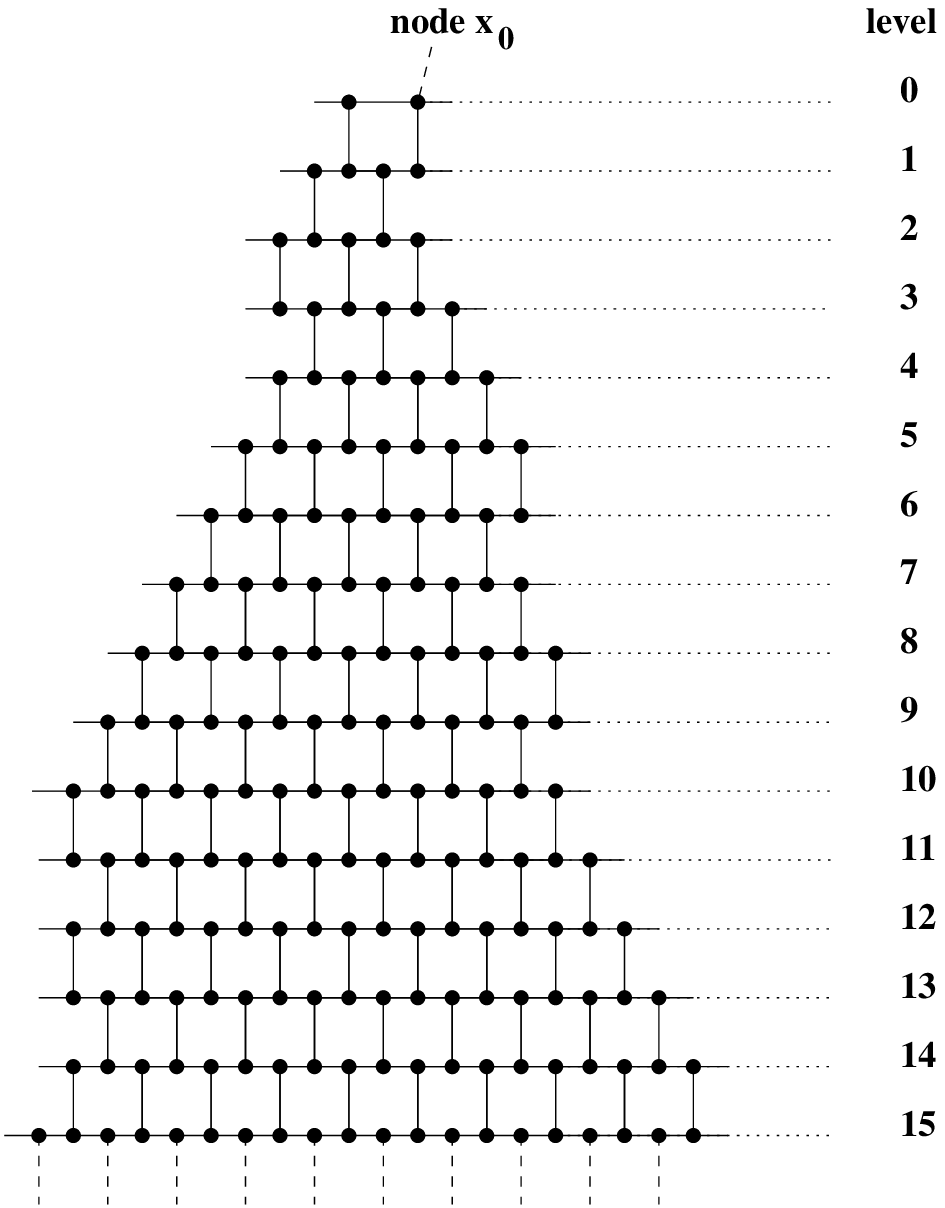}{tb}
Let $1 \leq D \leq 2$ be an arbitrary real number. Now we construct
the graph like in figure \ref{figbanana}. On level $m$ we use a width
of $\lfloor (2m - 1)^{D-1}\rfloor$ boxes. The construction is
continued ``downwards'' to infinity. To calculate the dimension we
observe that starting from $x_0$ we reach level $m$ after $n= 2m - 1$
steps. Thus we get with $n_k := 2k - 1$
\begin{align}
|\partial \U_{n_k}(x_0)| = \lfloor n_k^{D-1} \rfloor \thus \lim_{k
  \rightarrow  \infty} \frac{\ln|\partial \U_{n_k}(x_0)|}{\ln(n_k)}=
  D-1 \punkt
\end{align}
Using lemmas \ref{lem1}, \ref{lem2} and \ref{lem3} we see that this
graph has internal scaling dimension $D_S=D$.
If we close the construction horizontally, i.e.\ introduce bonds between
the leftmost and the rightmost nodes on each level we even can achieve
a completely homogeneous node degree $d=3$.

\begin{remark}
\begin{enumerate}
\item The constructed graph has privileged nodes, the one we
denoted as node $x_0$ and its counterpart on the same level.
\item Locally the constructed conical graph is completely isomorphic to
  a two-dimensional lattice. The non-integer dimension is
  only implemented as a global property of the graph.
\end{enumerate} 
\end{remark}

\subsection{Self-Similar Graphs} \label{sec4.2}
It is well known in graph theory that it is notoriously difficult to
construct large graphs with prescribed properties. It also proved
quite difficult to construct graphs with a prescribed (internal
scaling) dimension $D_S=D$ which don't exhibit the disadvantages of
the conical graphs described above. The main idea which solves the
problem is to use the well known theory of self similar sets or fractals
and their dimension theory. In the following we want to show
how this works and that we indeed can construct adjoint graphs to
self similar sets which have internal scaling dimension equal to the
box counting dimension of the self similar sets.

Given a strictly self similar set in $\R^p$ we canonically
construct an adjoint graph which also will be called
self-similar. 
The construction principle is based on an algorithm to compute the box
counting dimension of a self-similar set. We will illustrate our
proceedings with one main example. We construct a self-similar set
generated with the open unit square in $\R^2$ with lower left corner
at the origin and the similarity transforms
\begin{align}
  & S_1: ~\vec{x} \longmapsto \frac{1}{3} \,\vec{x} + \binom{0}{0} \komma
  \quad S_2: ~\vec{x} \longmapsto \frac{1}{3} \,\vec{x} +
  \binom{0}{\frac{2}{3}} \komma \quad
  S_3: ~\vec{x} \longmapsto \frac{1}{3} \,\vec{x} + \binom{\frac{1}{3}}
  {\frac{1}{3}} \\
  & S_4: ~\vec{x} \longmapsto \frac{1}{3} \,\vec{x} +
  \binom{\frac{2}{3}}{0} \komma \quad S_5: ~\vec{x} \longmapsto
  \frac{1}{3} \,\vec{x} + \binom{\frac{2}{3}}{\frac{2}{3}} \punkt
\end{align}
This set is sometimes called {\em Maltese Cross}, cf. \cite{7}.
The first construction steps are shown in figure \ref{figfracconstr}. For
details concerning self-similar sets and dimensions of fractals see
\cite{6}.
\fig{Construction steps of the example self-similar set}{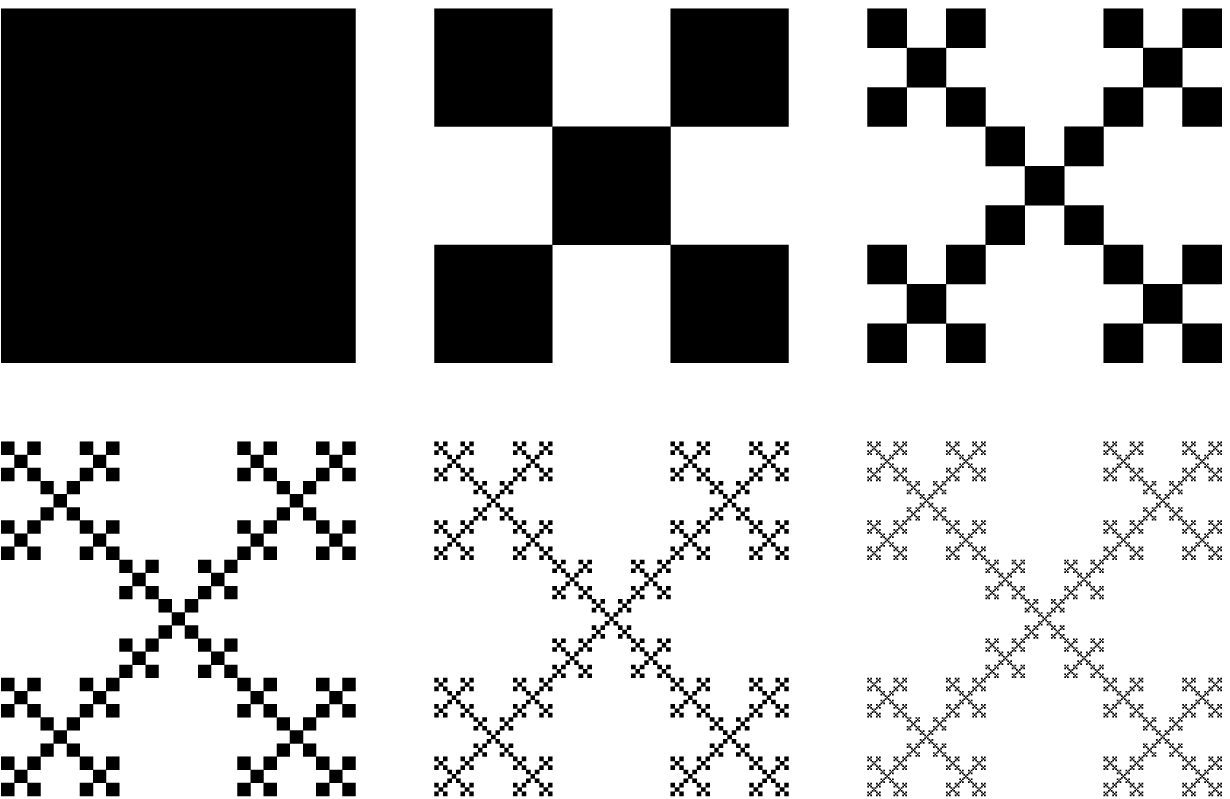}{tbp}

\subsubsection{Construction Based on Self-Similar Sets} \label{sec4.2.1}
Let $M$ be a strictly self-similar set with similarity transforms
$S_i$, $i \in I$, $I \subset \N$ and $|I| < \infty$. The contraction
factors $c_i$ of $S_i$ may all be equal, $c_i = c \in (0,1)$.
Now we cover $M$ with cubic lattices $L_n \subset \R^p$ with closed cubes of
edge length $c^n$, $n \in \N$, and replace every cube which has non-void
intersection with $M$ by a node. Nodes will be connected iff the
corresponding cubes in the covering cubic lattices have a non-void
intersection, i.e.~have a common corner or edge.

By this construction we get a finite graph $\G_n$ for
each $n \in \N$. The degree of these $\G_n$ is uniformly bounded
because an $n$-dimensional cube can only touch a finite number of
neighbor cubes in the cubic lattice. 
The graph we are interested in is $\G_\infty$, the graph we get through
infinite continuation of our construction. The first steps of this construction
scheme for our example are shown in figure \ref{figconstruct0}. 
\fig{Construction of graphs from self-similar sets}{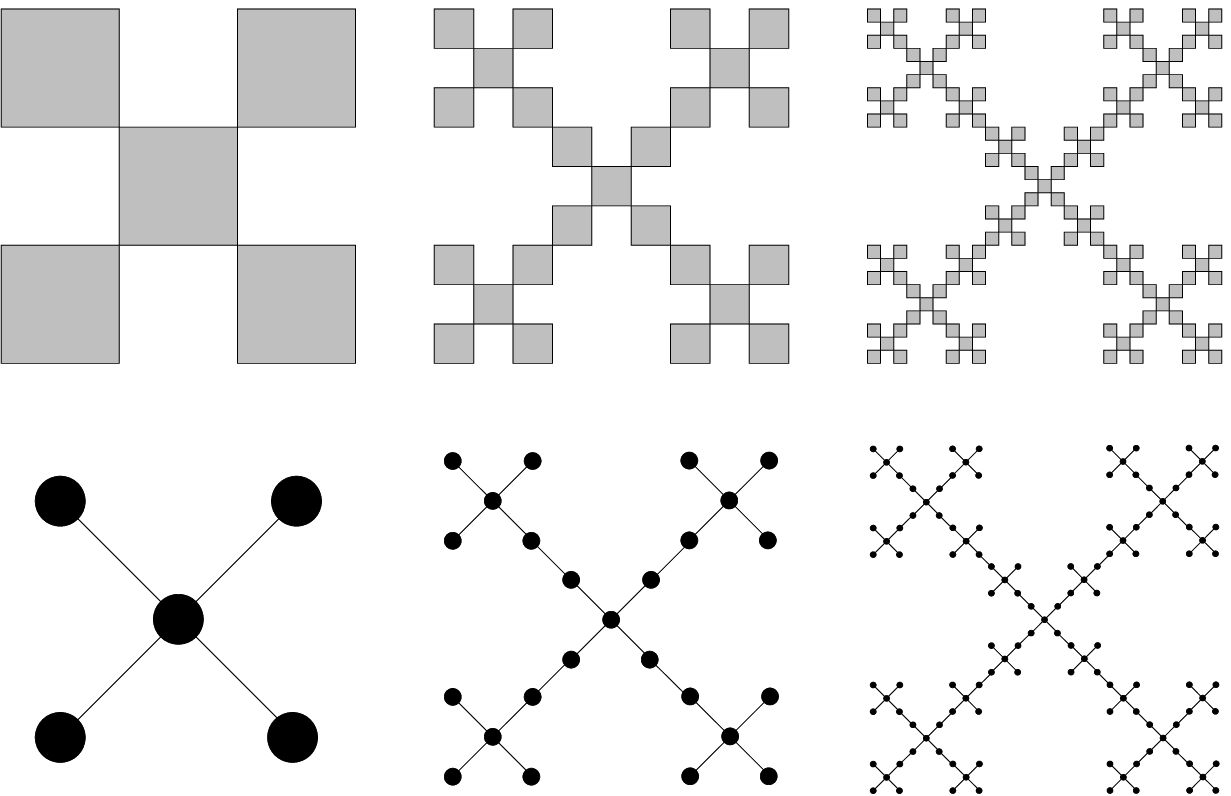}{tbp}
\begin{remark}
\begin{enumerate}
\item We will see later on, that no problems arise from the infinite
  continuation of the construction steps.
\item The self-similarity of $M$ transfers to \G ~in the sense that we
  can also define an equivalent of the similarity transforms of the
  self-similar set $M$. Details will become clear when we give a
  self-contained algorithm for the construction of self-similar graphs.
\item Connected self-similar sets produce connected self-similar
  graphs. The inverse is not true in general as our example shows.
  Here \G ~is connected but the self similar set we started with is
  not.
\end{enumerate}
\end{remark}

\subsubsection{Self-Contained Construction Algorithm} \label{sec4.2.2}
We want to illustrate two different views of a self-contained construction
algorithm for self-similar or hierarchical graphs.
\fig{Self-contained construction}{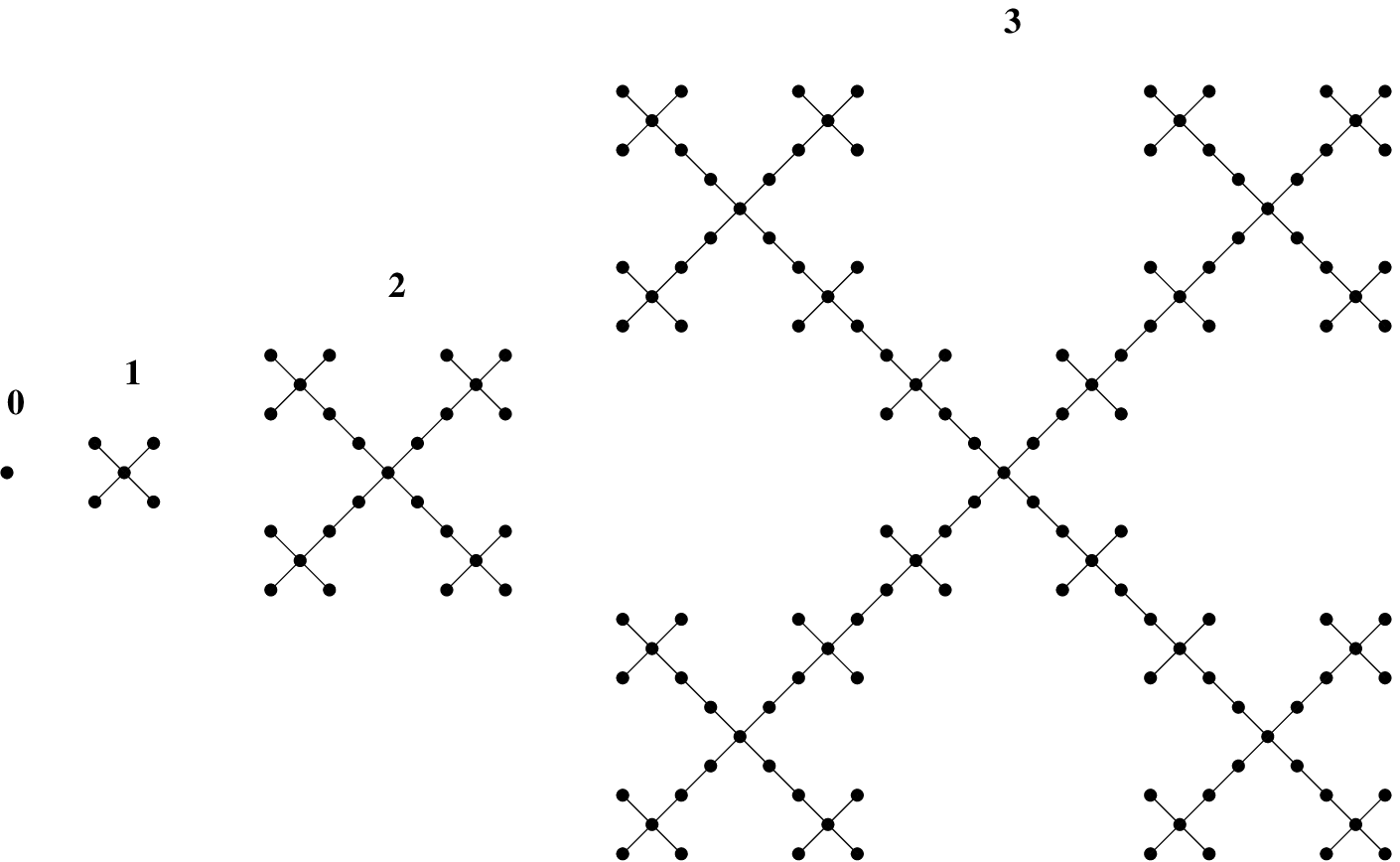}{tbp}
\begin{enumerate}
\item Construction by insertion: \label{enum1}
\begin{enumerate}
\item We start with a single node, $\G_0= (\{n_0\}, \emptyset)$.
\item $\G_1$ is the so-called generator, some finite graph. We denote the 
number of nodes in $\G_1$ as $N_g$.
\item We construct $\G_{n+1}$ from $\G_n$ by replacing every node in $\G_n$
by the generator $\G_1$ and interpret the original bonds in $\G_n$ as bonds
between some ``marginal'' nodes of the different copies of $\G_1$. In 
figure \ref{figconstruct1} we have drawn the first construction steps
of our example.
\end{enumerate}
\item Construction by ``copy and paste'': 
\begin{enumerate}
\item and~\addtocounter{enumii}{1}(\theenumii) are identical to \ref{enum1}.
\item We construct $\G_{n+1}$ from $\G_n$ by copying $\G_n$ $N_g$ times and
pasting these copies together in the same fashion as the nodes of the
generator are arranged. The construction steps can't be distinguished
from those in figure \ref{figconstruct1}.
\end{enumerate}
\end{enumerate}

\begin{remark} 
\begin{enumerate}
\item It becomes clear when looking at examples that the above construction
algorithms are equivalent.
\item The construction is -- of course -- not unique. The result strongly
depends on the choice of the nodes in $\G_{n+1}$ which carry the bonds
of $\G_n$ in the first construction or $\G_1$ in the second one respectively.
In our example all ``marginal'' nodes of the generator are equivalent because 
of the symmetry of the generator and therefore the construction is unique.
\item Seen from the viewpoint of the second construction it becomes
clear that the local neighborhood of any node doesn't change in the course
of the further construction. Therefore we can investigate any property
of \G ~in some $\G_N$ with sufficiently large $N$. Thus the infinite
continuation of construction steps needn't worry us at all.
\item The first construction scheme provides us with the analogon
of the simi\-lar\-ity transforms of the self-similar set. These transforms
correspond to the mapping of \G ~on $\tilde{\G}$ where $\tilde{\G}$ is
formed from \G ~like some $\G_{n+1}$ from $\G_n$. Clearly \G ~is invariant
under this mapping.   
\end{enumerate}
\end{remark}
As we can see from our example, all three construction algorithms, the 
self-contained ones as well as the one based on a self-similar set, are
equivalent provided the self-similar set and the choice of the generator
match. Seen in this light we can use all the construction principles
simultaneously in our arguments.

\subsubsection{Dimension of Self-Similar Graphs} \label{sec4.2.3}
Now we calculate the dimension of the graphs we get by the above
construction using some self-similar set $M$. For the sake of simplicity 
we assume that $\G_1$ has a central node $x_0$ in the sense that all
``marginal'' nodes which carry the ``outer'' bonds have all the same 
distance $r$ to this node. We further assume that $\frac{1}{c}$ ($c$ the
contraction parameter) is a natural number which is true in
most of the well known examples of self-similar sets
 and finally that the self-similar set produces a connected adjoint graph.
Then it is easy to see that starting from node $x_0$ we can exactly reach 
all nodes of construction step $k+1$
after $n_{k+1}= r + 2r\,n_k + n_k = (2r+1) \,n_k + r$ steps
in the graph, with - of course - $n_0 = 0$.
Thus $|\U_{n_k}(x_0)|$ is equal to the number of nodes in construction step
$k$, i.e. $|\U_{n_k}(x_0)| = N_{\delta_k} =
N_{c^k}$.\footnote{$N_{\delta_k}$ 
  is the number of cubes of edge length $\delta_k$ intersecting M,
  see the calculation of the box counting dimension in e.g. \cite{6}.}
Explicitly we get for $n_k$
\begin{align}
n_k = \sum_{j=0}^{k-1} (2r+1)^{j} \,r = r \,\frac{(2r+1)^k - 1}{2r} \quad
\forall k \geq 1 
\end{align}
Now let us relate $r$
to the contraction parameter $c$ of the self-similar set. We assumed
that the graph constructed from the self-similar set is
connected. This implies that there are $\frac{1}{c}$ nodes on the
``diagonal'' of the generator, i.e. $2r + 1 = \frac{1}{c}$.
Now we have for the internal scaling dimension of \G
\begin{align}
\lim_{k \rightarrow \infty} D_{n_k}(x_0) & = \lim_{k \rightarrow \infty}
\frac{\ln(N_{c^k})}{\ln\left(r\frac{(2r+1)^k-1}{2r}\right)} \\ 
& =  \lim_{k \rightarrow \infty} \frac{\ln(N_{c^k})}{\ln((2r+1)^k) + \ln\left(
\frac{1-(2r+1)^{-k}}{2r}\right)} \\
& = \lim_{k \rightarrow \infty} \frac{\ln(N_{c^k})}{-\ln(c^k) +
\ln\left(\frac{1-(2r+1)^{-k}}{2r}\right)} = \dim_B(M) 
\end{align}
in which $\dim_B(M)$ is the box counting dimension of $M$.
Of course lemmas \ref{lem2} and \ref{lem3} provide us with the knowledge
that this is the dimension of \G ~starting from any node.

Thus we established equality of the box counting dimension of self-similar
sets and the internal scaling dimension of the adjoint self-similar 
graphs under the assumptions stated above.
\begin{remark}
The assumed existence of a central node $x_0$ is not essential for the
equality of the dimensions of the fractal and the graph. The equality 
still holds in a more general context, e.g. for fractals like the 
Sirpinski Triangle. It is difficult though to give a general proof for
arbitrary self-similar sets.
\end{remark}

\subsubsection{Approximation of a Two Dimensional Lattice} \label{sec4.2.4}
In this paragraph we want to show how it now becomes possible to do
a dimensional approximation of a $n$-dimensional cubic lattice.
Again, for the sake of simplicity, we discuss the idea only 
with a two-dimensional lattice but the generalization to $n$ dimensions 
is obvious.

\fig{Some generators}{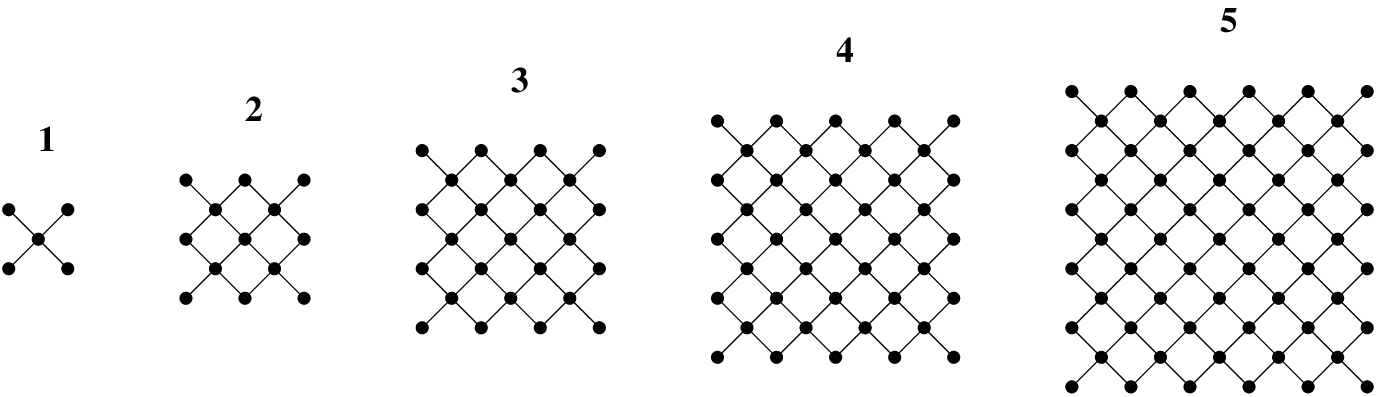}{tbp}
We introduce generators as shown in figure \ref{figgener}. With these we
get graphs of dimensions
\begin{align}
D_S^{(l)} = \frac{\ln(2 l^2 + 2 l +1)}{\ln(2 l +1)} 
\end{align}
in which $l$ is the number which labels the generators in figure
\ref{figgener}.
Obviously we have
\begin{align}
\lim_{l \rightarrow \infty} D_S^{(l)} = \lim_{l \rightarrow \infty}
\frac{\ln(2 l^2 + 2 l +1)}{\ln(2 l +1)} = \lim_{l \rightarrow \infty}
\frac{2\ln(l)+\ln(2+\frac{2}{l} + \frac{1}{l^2})}{\ln(l) +
  \ln(2+\frac{1}{l})}
= 2 \punkt
\end{align}
In this sense we have a dimensional approximation of a
two-dimensional lattice as alleged above. 
This might have some relevance in
connection with the dimensional regularization used in many
renormalization approaches to quantum field theory.
\begin{remark}
The generators above correspond to fractal sets known as ``sponges'',
see e.g. \cite{7}.
We can construct such ``sponges'' for any dimension $n$, we just need
to modify the generators appropriately. 
\end{remark}

\subsubsection{How to Change the Dimension of a Graph} \label{sec4.2.5}
To enlarge the dimension of a graph it is necessary to add either 
bonds or nodes to the graph. In the former case we
showed that adding only bonds between nodes with
original distance less than some $k \in \N$ does not change the
dimension. We want to illustrate this with an example. Let us try to
get a two-dimensional lattice starting from an one-dimensional one.
The procedure is shown in figure \ref{figsnail}. The dotted bonds are
those we added. As is easily seen, the former distance between the newly
connected nodes grows unboundedly with $n$, the number of the nodes in
the original graph. 
\fig{Deforming a one-dimensional graph into a two-dimensional one}{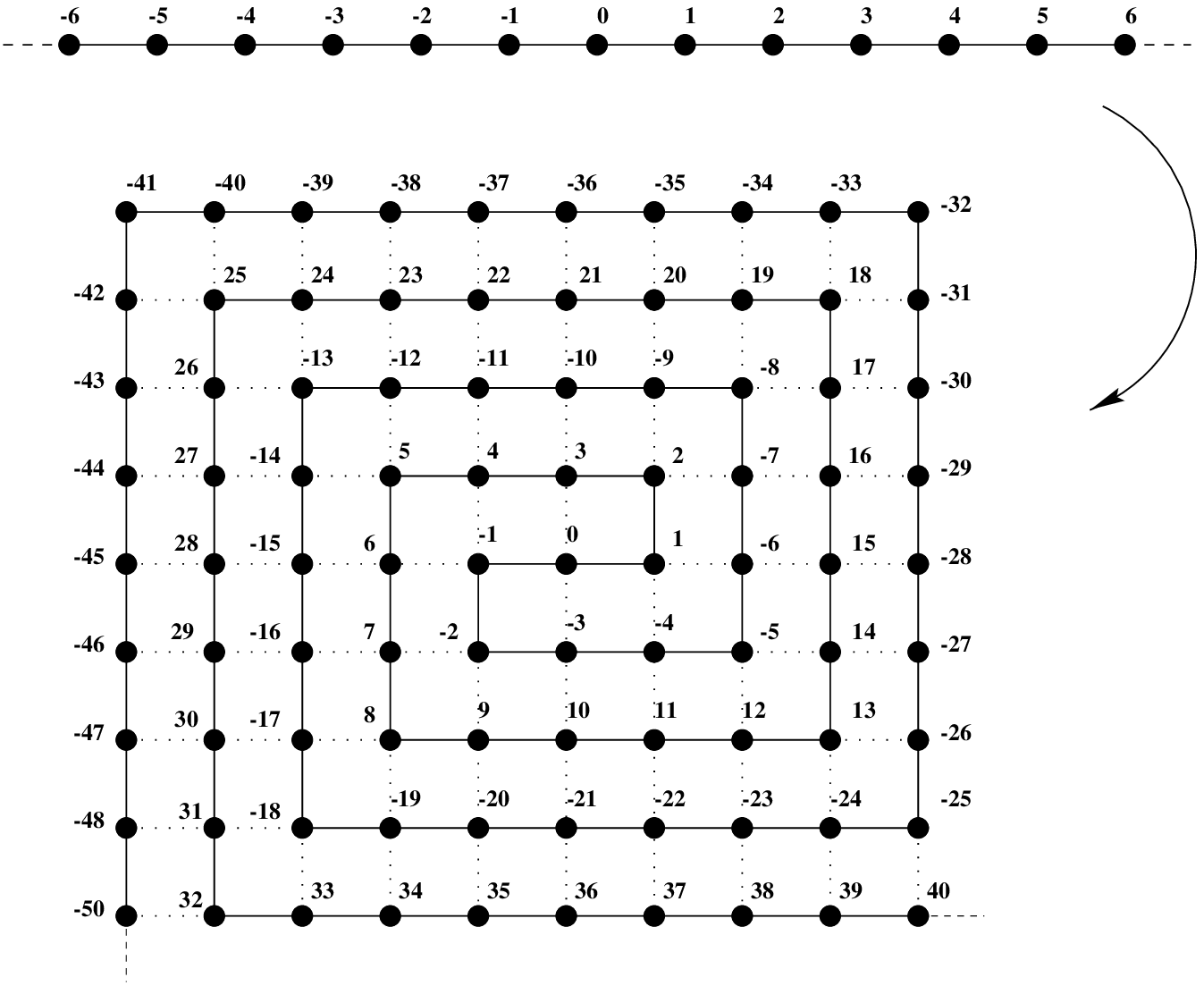}{tb}

If we choose to add nodes instead, it is equivalent to adding bonds
to new nodes which formerly had infinite distance to the nodes of the
original graph. This also illustrates the general result because
adding finitely many nodes certainly doesn't change the dimension.

%\section{Conclusions} \label{sec5}

\end{document}